\documentstyle[12pt]{article}
\newlength{\pubnumber} 

\catcode`\@=11
\@addtoreset{equation}{section}
  
\def\section{\@startsection{section}{1}{\z@}{3.5ex plus 1ex minus .2ex}
 {2.3ex plus .2ex}{\large\bf}}
\def\subsection{\@startsection{subsection}{2}{\z@}{2.3ex plus .2ex}
 {2.3ex plus .2ex}{\bf}}

\def\l{\langle}
\def\r{\rangle}
\def\vsq#1{\vert\l{#1}\r\vert^2}
\def\p23{\vsq{{\bar\Phi}_{23}}}
\def\v32{\vsq{V_3}}
\def\h18{\vsq{H_{18}}}
\def\anomaly{{{g^2}\over{16\pi^2}}{1\over{2\alpha^\prime}}}
\begin{document}

\begin{titlepage}
\samepage{
\setcounter{page}{1}
\rightline{OUTP--03--15P}
\rightline{\tt hep-ph/0306186}
\rightline{June 2003}
\vfill
\begin{center}
 {\Large \bf  String Inspired Neutrino Mass Textures \\
in Light of KamLAND and WMAP}
\vfill
 {\large
Claudio Coriano$^1$\footnote{E-mail address: Claudio.Coriano@le.infn.it}
 $\,$and$\,$
Alon E. Faraggi$^2$\footnote{E-mail address: faraggi@thphys.ox.ac.uk}
   \\}
\vspace{.12in}
{\it $^1$Dipartimento di Fisica,
 Universita' di Lecce,\\
 I.N.F.N. Sezione di Lecce,
Via Arnesano, 73100 Lecce, Italy\\}
\vspace{.05in}
 {\it $^{2}$   Theoretical Physics Department, 
		University of Oxford,
		Oxford OX1 3NP UK \\}
\end{center}
\vfill
\begin{abstract}
  {\rm
Recent data from astrophysical and terrestrial experiments
indicates large mixing angles in the neutral lepton sector and
restricts the allowed regions of neutrino masses. In particular, 
the large mixing angles in the lepton sector are disparate from 
the small mixing in the quark sector. This disparity is unnatural from
the point of view of grand unified theories, that are well motivated
by the Standard Model multiplet structure and logarithmic running
of its parameters. we examine the issue of this disparity from 
the perspective of string derived $SO(10)$ GUT models, in which
the $SO(10)$ symmetry is broken directly at the string theory level. 
A characteristic feature of such models is the appearance of
numerous $SO(10)$ singlet fields. We propose that the mismatch
between the quark and lepton mixing parameters arises due to
this extended singlet spectrum and its mixing with the right--handed
neutrinos. We discuss a string inspired effective parameterization of the
extended neutrino mass matrix and demonstrates that the
coupling with the $SO(10)$ singlet spectrum can readily account
for the neutrino flavor parameters. The mechanism implies that some
$SO(10)$ singlet fields should exist at intermediate mass scales. 
We study the possibilty of deriving the neutrino mass
spectrum from string theory in a specific string derived 
vacuum solution, and comment on the properties
that such a solution should possess.

}
\end{abstract}
\vfill
\smallskip}
\end{titlepage}

\setcounter{footnote}{0}

\def\beq{\begin{equation}}
\def\eeq{\end{equation}}
\def\beqn{\begin{eqnarray}}
\def\eeqn{\end{eqnarray}}
\def\AEF{A.E. Faraggi}
\def\NPB#1#2#3{{\it Nucl.\ Phys.}\/ {\bf B#1} (#2) #3}
\def\PLB#1#2#3{{\it Phys.\ Lett.}\/ {\bf B#1} (#2) #3}
\def\PRD#1#2#3{{\it Phys.\ Rev.}\/ {\bf D#1} (#2) #3}
\def\PRL#1#2#3{{\it Phys.\ Rev.\ Lett.}\/ {\bf #1} (#2) #3}
\def\PRT#1#2#3{{\it Phys.\ Rep.}\/ {\bf#1} (#2) #3}
\def\MODA#1#2#3{{\it Mod.\ Phys.\ Lett.}\/ {\bf A#1} (#2) #3}
\def\IJMP#1#2#3{{\it Int.\ J.\ Mod.\ Phys.}\/ {\bf A#1} (#2) #3}
\def\nuvc#1#2#3{{\it Nuovo Cimento}\/ {\bf #1A} (#2) #3}
\def\JHEP#1#2#3{ {\it JHEP } {\bf #1} (#2)  #3}
\def\etal{{\it et al,\/}\ }
\hyphenation{su-per-sym-met-ric non-su-per-sym-met-ric}
\hyphenation{space-time-super-sym-met-ric}
\hyphenation{mod-u-lar mod-u-lar--in-var-i-ant}


\setcounter{footnote}{0}
\section{Introduction}

The neutrino sector of the Standard Model provides another piece
to the flavor enigma. Evidence for neutrino oscillations
steadily accummulated over the past few years, resulting in
compelling evidence for neutrino masses. This in turn points
to the augmentation of the Standard Model by the right--handed
neutrinos, and provides further evidence for the elegant
embedding of the Standard Model matter states, generation by generation,
in the 16 spinorial representation of $SO(10)$. However, in this
respect the new neutrino data raises further puzzles. The 
observation of a zenith angle dependence of $\nu_\mu$ 
from cosmic ray showers at super-Kamiokande \cite{superk}
provides strong evidence for oscillations in atmospheric neutrinos
with maximal $\nu_\mu\rightarrow\nu_\tau$ oscillations,
whereas the observations at the solar Sudbury Neutrino Observatory
(SNO) \cite{sno} and at the reactor KamLAND experiment \cite{kamland}
favor the large mixing angle MSW solution of the solar neutrino problem
\cite{MSW}. The recent data from the Wilkinson Microwave Anisotropy
Probe (WMAP) on cosmic microwave background anisotropies \cite{wmap},
combined with the 2 degree Field Galaxy Redshift Survey, CBI and ACBAR
\cite{galaxy},  restritcs the amount of critical density 
attributed to relativistic neutrinos, and imposes that the
sum of the masses is smaller than $0.75$eV. 

While the Standard Model data strongly supports the incorporation of the 
Standard Model gauge and matter spectrum in representations
of larger gauge groups,
the flavor sector of the Standard Model provides further 
challenges. In the heavy generation the consistency
of the bottom--quark--tau lepton mass ratio with the 
experimental data arises due to the 
running of the strong gauge coupling.
The remaining flavor data, however, must
have its origin in a theory that incorporates gravity
into the picture. Most developed in this context are
the string theories that provide a viable perturbative framework
for quantum gravity. However, a new twist
of the puzzle arises due to the fact that while in
the quark sector we observe an hierarchical mass pattern
with suppressed mixing angles, the observations in the
neurtino sector are compatible with large mixing angles
that implies approximate mass degeneracy.

An elegant mechanism in the context of $SO(10)$ unification
to explain the large mixing in the neutrino sector was proposed
in ref. \cite{bajc}. However, this mechanism utilizes the 126 of
$SO(10)$, that does not arise in perturbative string theories \cite{dmr}.
On the other hand, string constructions
offer a solution to the proton longevity
problem. A doublet--triplet splitting mechanism is induced
when the $SO(10)$ symmetry is broken to $SO(6)\times SO(4)$
by Wilson--lines \cite{ps}. In the stringy doublet--triplet splitting
mechanism the color triplets are projected from the massless spectrum
and the doublets remain light. Additional symmetries that arise in the
string models may also explain the suppression of proton decay from
dimension four and gravity mediated operators. String constructions
also explain the existence of three generations in terms of the 
geometry of the compactified manifold. It is therefore 
important to seek other explanations for the origin of the 
discrepancy in the quark and lepton mass sectors.
An alternative possibility to the utilization of the 
126 in the seesaw mechanism is to use the nonrenormalizable term
$1616{\overline{16}}{\overline{16}}$. In this case the $B-L$ symmetry
is broken along a supersymmetric flat direction by the VEVs
of the neutral components of
$\langle16_H\rangle=\langle{\overline{16}}_H\rangle$,
where $16_H$ and ${\overline{16}}_H$ are two Higgs
multiplets, distinct from the three Standard Model generations. 
This term then induces the heavy Majorana mass term for the
right--handed neutrino. The contemporary studies of neutrino
masses in this context are based on this term. We will refer to this
as the ``one--step seesaw mechanism'' \cite{seesaw}. Similarly,
explorations in the context of type I string inspired models
also use the ``one--step seesaw mechanism'' \cite{king}.
In this paper we propose that the neutrino data points to the
role of $SO(10)$ singlet fields in the see--saw mass matrix.

\section{Summary of neutrino data}

In this section we summarize the neutrino data.
The KamLAND  and SNO data are compatibe with the large mixing angle
solution to the solar neurtino puzzle, with
\beq 
\Delta m_{\rm 12}\approx7.1\times 10^{-5}{\rm eV}^2;~
\sin^22\theta_{\rm 12}\ge0.86
\eeq

The super-Kamiokande gives

\beq 
\Delta m_{\rm 23}\approx2.7\times 10^{-3}{\rm eV}^2;~
\sin^22\theta_{\rm 23}\approx1.00
\eeq
The $\theta_{13}$ angle is contrained by the CHOOZ experiment \cite{chooz}
with, $\sin\theta_{13}\le0.2$.

Assuming that all the light neutrinos are stable
the WMAP data yields an upper bound on the sum of
neutrino masses
\beq
\sum_i m_i < 0.71{\rm eV}.
\label{wmap}
\eeq

We assume here a $SO(10)$ symmetry that underlies the Standard Model
spectrum and interactions. This underlying symmetry may be broken
directly at the string scale and need not be present in the effective
low energy field theory. However, it indicates that the lepton
mass matrices are related to the quark mass matrices. In particular,
the mixing angles in the charged--lepton sector and the
Dirac neutrino mass matrix are related to the corresponding angles
in the quark sector, which are small. With the assumption we
can take the charged--lepton mass matrix to be diagonal,
in which case the mixing information is contained entirely
in the Majorana neutrino mass matrix.

In the mass basis the neutrino mass matrix, $M_D$, is
diagonal, and is related to the neutrino mass matrix
in the flavor basis by a unitary transformation
\beq
M_\nu=U M_D U^T
\label{mnumd}
\eeq
The MNS mixing matrix $U$ relates between the flavor and mass eigenstates
\beq
{\left(\matrix{
                 {\nu_e}\cr{\nu_\mu}\cr{\nu_\tau}\cr
                }
   \right)}=
  {\left(\matrix{
                 U_{e1}  &U_{e2}  & U_{e3}\cr
                 U_{\mu1}&U_{\mu2}& U_{e3}\cr
                  U_{\tau1}&U_{\tau2}& U_{\tau3}\cr
                }
   \right)}
  {\left(\matrix{
                 {\nu_1}  \cr
                 {\nu_2}\cr
                 {\nu_3} \cr
                }
   \right)}.
\eeq
Assuming no CP violation $U$ can be written as
$U=U_{23}U_{13}U_{12}$, where $U_{ij}$ related between
the $i,j$ mass eigenstates. The elements of $U_{ij}$ are constrained
by the atmospheric, reactor, and solar neutrino data,
and one can then obtain a simple form for the
mixing matrix $U$. Given the relation (\ref{mnumd}),
and the experimental constraints on the neutrino mass
differences, the allowed patterns of the Majorana
neutrino mass matrices, $M_\nu$, can then be classified
\cite{mnuclasses}. These are the mass matrices that 
one would like to obtain from string theory.

\section{General structure of the string models}

In this section we discuss the general structure of string models
and the new features that arise from them. 
A class of semi--realistic string models that preserve the $SO(10)$
embedding of the Standard Model spectrum are the heterotic--string
models in the free fermionic formulations
\cite{revamp,slm,alr,eu,ks,chl,lrs,su421}.
While quasi--realistic string models that contains three generations
with the correct charge assignment under the Standard Model gauge
group are quite abundant, the free fermionic models
preserve the $SO(10)$ embedding. This class of models serves as the
prototype laboratory for phenomenological exploration of
$SO(10)$ string GUTs.

In this section we summarize the general structure of the realistic
free fermionic models, and of their massless spectra. 
The free fermionic heterotic--string formulation yields a large number 
of three generation models, which possess an underlying $Z_2\times Z_2$
orbifold structure, and differ in their detailed phenomenological 
characteristics. We discuss here 
the common features of this 
large class of realistic string models. The discussion is qualititative
and details are given in the references cited. In section 
(\ref{278}) we analyze one specific string model in more detail.

The free fermionic models are constructed by specifying a set
of boundary conditions basis vectors and the one--loop
GSO projection coefficients \cite{fff}.
The basis vectors, $b_k$, span a finite  
additive group $\Xi=\sum_k{{n_k}{b_k}}$.
The physical massless states in the Hilbert space of a given sector
$\alpha\in{\Xi}$, are obtained by acting on the vacuum with 
bosonic and fermionic operators and by
applying the generalized GSO projections.

The four dimensional gauge group in the three generation
free fermionic models arises as follows. The models can 
in general be regarded as constructed in two stages.
The first stage consists of the NAHE set of boundary conditions basis
vectors, which is a set of five boundary condition basis vectors, 
$\{{\bf1},S,b_1,b_2,b_3\}$ \cite{nahe}. 
The gauge group after imposing the GSO projections induced
by the NAHE set basis vectors is $SO(10)\times SO(6)^3\times E_8$
with $N=1$ supersymmetry. The sectors $b_1$, $b_2$ and $b_3$
produce 48 multiplets in the $16$ representation of $SO(10)$, that are
are singlets of the hidden $E_8$ gauge group, and transform 
under the horizontal $SO(6)_j$ $(j=1,2,3)$ symmetries.
The untwisted sector produces states in the 10 vectorial representation
of $SO(10)$, that can produce electroweak Higgs doublets,
and $SO(10)$ singlets that are charged under the $SO(6)^3$ symmetries.
This structure
is common to all the realistic free fermionic models. At this stage 
we anticipate that the $SO(10)$ group gives rise to the Standard Model 
group factors, {\it i.e.} to the $SO(10)$ GUT symmetry.
The $161610$ $SO(10)$--invariant coupling can then gives rise to the
Dirac fermion mass terms.

The second stage of the free fermionic
basis construction consists of adding to the 
NAHE set three (or four) additional boundary condition basis vectors,
typically denoted by $\{\alpha,\beta,\gamma\}$.
These additional basis vectors reduce the number of generations
to three chiral generations, one from each of the sectors $b_1$,
$b_2$ and $b_3$, and simultaneously break the four dimensional
gauge group. The $SO(10)$ symmetry is broken to one of its subgroups
$SU(5)\times U(1)$ (FSU5) \cite{revamp}, $SO(6)\times SO(4)$ (PS)
\cite{alr},
$SU(3)\times SU(2)^2\times U(1)$ (SLM) \cite{slm},
$SU(3)\times SU(2)\times U(1)^2$ (LRS) \cite{lrs}, or
$SU(4)\times SU(2)\times U(1)$ (SU421) \cite{su421}.
Similarly, the hidden $E_8$ symmetry is broken to one of its subgroups.
The basis vectors $\{\alpha, \beta, \gamma\}$, combined with the
NAHE--set basis vectors, give rise to additional massless sectors. 
In the FSU5 and PS type models two of these sectors produce 
the GUT Higgs representations that break the GUT symmetry,
that typically appear as $16_H\oplus\overline{16}_H$
decomposed under the final $SO(10)$ unbroken subgroup.
This states can therefore be used to induce the heavy Majorana
mass scale for the right--handed neurtino \cite{lola}, 
in a ``one--step seesaw'' mechanism.
Additionally, the models contain massless states from the sectors
that break the $SO(10)$ symmetry. These states cannot be embedded
in $SO(10)$ representations, and carry fractional charge under either
the electroweak hypercharge, or under the $SO(10)$
$U(1)_{Z^\prime}$--subgroup \cite{ccf}.
In the SLM models of ref. \cite{slm,eu}
the neutral component of $\overline{16}_H$
does not arise in the spectrum. Instead, the models contain
Standard Model singlet states that carry 1/2 of the $U(1)_{Z^\prime}$
charge of the righ--handed neutrino. These states are utilized
in the implimentation of the seesaw mechanism in the free fermionic
standard--like models \cite{fh,hnm}.

Subsequent to constructing the basis vectors and extracting the massless
spectrum the analysis of the free fermionic models proceeds by
calculating the superpotential. The cubic and higher-order terms in
the superpotential are obtained by evaluating the correlators
\beq
A_N\sim \langle V_1^fV_2^fV_3^b\cdots V_N\rangle,
\label{supterms}
\eeq
where $V_i^f$ $(V_i^b)$ are the fermionic (scalar) components
of the vertex operators, using the rules given in~\cite{kln}.
Generically, correlators of the form (\ref{supterms}) are of order
${\cal O} (g^{N-2})$, and hence of progressively higher orders
in the weak-coupling limit. Typically, 
one of the $U(1)$ factors in the free-fermion models is anomalous,
and generates a Fayet--Ilioupolos term which breaks supersymmetry
at the Planck scale \cite{dsw}. A supersymmetric vacuum is obtained by
assiging non--trivial VEVs to a set of Standard Model singlet
fields in the massless string spectrum along $F$ and $D$--flat directions. 
Some of these fields will appear in the nonrenormalizable terms
(\ref{supterms}), leading to
effective operators of lower dimension. Their coefficients contain
factors of order ${\cal V} / M{\sim 1/10}$.

Pursuing this methodology the structure of the fermion
mass matrices in the free fermionic models was studied 
\cite{fermionmasses,ckm}, as well as in other string models
\cite{otherfermionmasses}.
The general texture of the quark mass matrices in the superstring 
standard--like models is of the following form \cite{ckm}, 
\beq
M_U\sim\left(\matrix{\epsilon,a,b\cr
                    {\tilde a},A,c \cr
                    {\tilde b},{\tilde c},\lambda_t\cr}\right);{\hskip .2cm}
M_D\sim\left(\matrix{\epsilon,d,e\cr
                    {\tilde d},B,f \cr
                    {\tilde e},{\tilde f},C\cr}\right);{\hskip .2cm}
\label{quarkmassmatrices}
\eeq
Due to the underlying $SO(10)$ symmetry structure
we anticipate the relations
\beq
M_E\sim M_D~~~~~~;~~~~~~M_N\sim M_U,
\label{quarkleptonrelation}
\eeq
where $M_E$ and $M_N$ are the charged--lepton and Dirac neutrino mass
matrices, respectively. In some models the top quark Yukawa coupling
$\lambda_t$ is
the only one that arises at the cubic level of the superpotential
and is of order one. The remaining quark and lepton Yukawa couplings
arise from higher order terms in the superpotential that are
suppressed relative to the leading cubic level term.

\section{String inspired neutrino mass textures}

In this section we discuss the string inspired two step seesaw mechanism,
and identify the features that the string models should produce to
accommodate this mechanism. The low energy effective field theory of our
string inspired model consist of three chiral $SO(10)$ generations
decomposed under the final unbroken $SO(10)$ subgroup. For concretness we
consider here the case of the standard--like models. In this case the 
unbroken $SO(10)$ subgroup is $SU(3)\times SU(2)\times
U(1)_{T_{3_R}}\times U(1)_{B-L}$. In addition to the three chiral
generations the matter spectrum contains two electroweak Higgs doublets;
the fields $N$ and $\bar N$ that break the additional $U(1)$ symmetry to
the Standard Model weak hypercharge. Additionally the model contains
three $SO(10)$ singlet fields $\phi_m$, that obtain an electroweak
or intermediate scale VEV.
The relevant terms in the superpotential that contribute to the
neutrino seesaw mass matrix are given by 
\beq
W=\cdots+\lambda_4^{ij}N_iL_j{\bar h}+
\lambda^{im}_5N_i{\bar N}{\phi}_m+\lambda_6^{ijk}\phi_i\phi_j\phi_k~.
\label{superpot}
\eeq
The first term produces the neutrino Dirac mass matrix, which by the
underlying $SO(10)$ is proportional to the up--quark mass matrix, $M_U$.
The second produces the $N\phi$ mass matrix, $M_\chi$. The terms in this
matrix are generated by VEVs that break the $B-L$ symmetry. However, 
contrary to the situtation in conventional GUTs, the $N\phi$ mass
terms can vary over several orders of magnitude. The reason
being that in the string models they are generically obtained from
nonrenormalizable operators that arise at different orders.
The third term Eq. (\ref{superpot})
produces the mass terms for the $SO(10)$ singlets.
Thus, the neutrino mass matrix at the unification scale
takes the general form \cite{rnm}
\begin{equation}
  {\left(\matrix{
                   0  &     M_D &   0       \cr
                 M_D  &     0   & M_{\chi}  \cr
                   0  &  M_\chi & M_\phi    \cr
                }
   \right)}
\label{nmm}
\end{equation}
and is the two--step seesaw mass matrix.
The left--handed Majorana neutrino mass matrix is given by
\beq
M_\nu=M_D M_\chi^{-1} M_\phi M_\chi^{-1} M_D^T~.
\label{mnu}
\eeq

Eq. (\ref{mnu}) has the following implications. First, it is noted
that it produces a double suppression with respect to the right--handed
neutrinos mass scales, which allows these to be intermediate
rather than at the GUT scale. This possibility is advantegeous
for the generation of the baryon asymmetry by leptogenesis 
\cite{leptogenesis}. Second, as discussed above, $M_D\sim M_U$
and can therefore be taken to be diagonal, {\it i.e.}
\beq
M_D\sim {\rm Diag}(m_u,m_c,m_t)
\label{diagmd}
\eeq
For simplicity we also take $M_\chi$ to be diagonal,
although one can consider the possibility that it is 
not. In this case, it is seen that the flavor structure of the 
left--handed neutrino mass matrix arises from the flavor 
structure of the matrix $\phi$. 

This result arises due to the extended singlet spectrum in the string
model that is external to the $SO(10)$ gauge group. To emphasize
this point it is instructive to write the see--saw mass matrix
in the form
\begin{equation}\label{svss}
{\bf M}=\left(\begin{array}{cc}0&{\bf H}\\{\bf H}^T&{\bf J}\end{array}\right).
\end{equation}
where
\begin{eqnarray}
{\bf H}=\left(\begin{array}{cccccccc}
m_u &  0  &  0  & 0 & 0 & 0   \\
 0  & m_c &  0  & 0 & 0 & 0   \\
 0  &  0  & m_t & 0 & 0 & 0    \end{array}\right)
\end{eqnarray}
and 
\begin{eqnarray}
{\bf J}=\left(\begin{array}{cccccccc} 
0 & 0 & 0 & \chi_1 &   0     & 0     \\
0 & 0 & 0 &   0    & \chi_2  & 0     \\
0 & 0 & 0 &   0	   &   0     &\chi_3 \\
\chi_1 & 0 & 0 & \phi_{11} & \phi_{12} & \phi_{13}\\
0 & \chi_2 & 0 & \phi_{21} & \phi_{22} & \phi_{23}\\
0 & 0 & \chi_3 & \phi_{31} & \phi_{32} & \phi_{33}\\
\end{array}\right),
\label{jmatrix}
\end{eqnarray}

In the limit of small left--handed neutrino masses 
the left--handed Majorana mass matrix is given by \cite{thor}
\begin{equation}
M_\nu \approx {\bf{H}}^T{\bf{J}}^{-1}{\bf{H}},   
\label{lightneutrinomassmatrix}
\end{equation}
and is identical to Eq. (\ref{mnu}). The left--handed Majorana
mass matrix $M_\nu$ is then given by
\begin{eqnarray}
{M_\nu}=
\left(\begin{array}{ccc}
{{m_u m_u}\over{\chi_1\chi_1}} \phi_{11}& 
{{m_u m_c}\over{\chi_1\chi_2}} \phi_{12}& 
{{m_u m_t}\over{\chi_1\chi_3}} \phi_{13}  \\
{{m_u m_c}\over{\chi_2\chi_1}} \phi_{21}& 
{{m_c m_c}\over{\chi_2\chi_2}} \phi_{22}& 
{{m_c m_t}\over{\chi_2\chi_3}} \phi_{23}  \\
{{m_u m_t}\over{\chi_3\chi_1}} \phi_{31}& 
{{m_c m_t}\over{\chi_3\chi_2}} \phi_{32}& 
{{m_t m_t}\over{\chi_3\chi_3}} \phi_{33}  \\
\end{array}\right),
\label{mnuleft}
\end{eqnarray}
The seesaw equation, eq. (\ref{lightneutrinomassmatrix})
on the other hand the 
same structure as the one--step seesaw mechanism. Namely, it has the
form $m^2/M$. This form demonstrates how the extended
singlet spectrum that couples to the right--handed neutrinos
results in the two--step seesaw mass matrix. We then note that
choosing the appropriate values for the parameters in $M_\chi$
and $M_\phi$ produces the desired left--handed neutrino mass
texture. For example, taking
$\chi_i=10^8{\rm GeV}$ and 
\begin{eqnarray}
{\bf M_\phi}=\left(\begin{array}{ccc}
10^{13} & 10^{10} & 10^8 \\
10^{10} & 10^{7}  & 10^5 \\
10^{8}  & 10^{5}  & 10^3 \\
\end{array}\right){\rm GeV},
\end{eqnarray}
produces a demotratic left--handed neutrino mass texture, with one
eigenvalue of order $1$eV. Obviousely, the additional freedom, in
principle, admits the required neutrino mass textures. Next we turn
to examine the neutrino mass textures in a specific $SO(10)$
string--GUT model.

\section{specific string model}\label{278}

As a concrete string model we examine the neutrino masses in 
the standard--like model of ref. \cite{eu}. 
The full massless spectrum of this model is given in
Ref. \cite{eu}. A partial set which is relevant for our
purposes includes the following states:

(I) There are three chiral families of quarks
and leptons, each with sixteen components, including ${\bar\nu}_R$,   
which arise from the twisted sectors $b_1$, $b_2$ and $b_3$.
These transform as 16's of $SO(10)$ and are neutral under $G_H$.

(II) the Neveu--Schwarz (NS) sector produces, in addition to the
gravity multiplets, three pairs of electroweak
doublets $\{h_1,h_2,h_3,{\bar h}_1,{\bar h}_2,{\bar h}_3\}$,
three pairs of $SO(10)$ -- singlets with $U(1)_i$ charge
$\{\Phi_{12},\Phi_{23},\Phi_{13},{\bar\Phi}_{12},{\bar\Phi}_{23},
{\bar\Phi}_{13}\}$, and three states that are singlets
of the entire four dimensional gauge group, $\{\xi_1,\xi_2,\xi_3\}$.

(III) the sector $S+b_1+b_2+\alpha+\beta$ produces one additional pair
of electroweak doublets $\{h_{45},{\bar h}_{45}\}$, one pair of color
triplets $\{D_{45}, {\bar D}_{45}\}$ and seven pairs of $SO(10)$
singlets with $U(1)_i$ charge $\{\Phi_{45},{\bar\Phi}_{45},
\Phi_{1,2,3}^\pm,{\bar\Phi}_{1,2,3}^\pm\}$.  

(IV) The sectors $b_j+2\gamma+
(I={\bf 1}+b_1+b_2+b_3)$ produce hidden--sector multiplets
$\{T_i,{\bar T}_i,V_i,{\bar V}_i\}_{i=1,,2,3}$ which are $SO(10)$
singlets but are non--neutral under $U(1)_i$ and the hidden $G_H$. 
The $T_i({\bar T}_i)$ are 5$(\bar5)$ and $V_i({\bar V}_i)$
are 3$(\bar3)$ of $SU(5)_H$ and $SU(3)_H$ gauge groups, respectively.

(V) The vectors in some combinations of
$(b_1,b_2,b_3,\alpha,\beta)\pm\gamma+(I)$ produce additional states 
which are either singlets of $SU(3)\times SU(2)\times U(1)_Y
\times SU(5)_H\times SU(3)_H$
or in vector--like representation of this group. The relevant states 
of this class $\{H_{17}-H_{26}\}$.
The states of class (V) are crucial for
the seesaw mechanism in the superstring standard--like models \cite{fh}.

One characteristic feature of this class of models,
is that, barring the three chiral 16's
there are no additional vector--like
$16+{\overline{16}}$ pairs. As a result, elementary fields with the
quantum numbers of $N_L'\in{\overline{16}}$ do not exist  
in this class of models.
Nevertheless, VEVs of products of certain condensates,
which are expected to form through the hidden sector
force and certain fields belonging to the set (V)
can provide the desired quantum numbers of sneutrino
like fields -- i.e ${\bar N}_R\in 16$ and ${N_L'}\in{\overline{16}}$,
as, for example, in the combinations shown below :
\begin{eqnarray}
&&\langle H_{19}{\bar T}_i\rangle\langle H_{23}\rangle\rightarrow
(B-L=-1,T_{3_R}=1/2)\sim N_L'\in {\overline{16}}\nonumber\\  
&&\langle H_{20}{T}_i\rangle\langle H_{26}\rangle\rightarrow
(B-L=+1,T_{3_R}=-1/2)\sim {\bar N}_R\in {{16}}
\label{effectiveN}
\end{eqnarray}
Note $H_{19}$ and $T_i$ transform as ${5}$, and $H_{20}$ and
$T_i$ transform as 5, of $SU(5)_H$, respectively.
Thus, $H_{19}(H_{20})$ can pair up with
${\bar T}_i(T_i)$ to make condensates
at the scale $\Lambda_H$, where $SU(5)_H$ force becomes strong.
In this model an effective
seesaw mechanism \cite{fh}
is implemented
by combining the familiar Dirac masses of the neutrinos which arise
through electroweak--symmetry breaking
scale, with superheavy mass terms which mix
${\bar \nu}_R^i$ with the singlet $\phi$ fields in sets
(II) and (III) \cite{fh}. 

The set of fields that enter the seesaw mass matrix includes
the three left--handed neutrinos, $L_i$; the three right--handed
neutrinos, $N_i$; and the set of Standard Model singlets. These
include: the $SO(10)$ singlets with $U(1)$ and hidden charges,
$\{\Phi_{45},\Phi_{1,2,3}^\pm,\Phi_{13},\Phi_{23},\Phi_{12},
T_i,V_i\}\oplus{\rm h.c.}$ The set of $SU(3)\times SU(2)\times U(1)_Y$
singlets with $U(1)_{Z^\prime}$ charge, $H_{13-14,17-20,23-26}$.
The set of entirely neutral singlets $\xi_{1,2,3}$. The analysis
now proceeds by analyzing the cubic level and higher order terms
in the superpotential. When shifting the vacuum by the anomalous
$U(1)$ cancelling VEVs, two competing processes, that depend
on the set of VEVs, take place. Some nonrenormalizable operators 
induce effective renormalizable operators. Obviously,
the number of unsuppressed operators increases
with the increasing number of fields with non--vanishing VEV.
At the same time, the number of Standard Model singlets that
receive heavy mass and decouple from the low energy spectrum
also increases. In fact, a priori it is not apparent that 
any of the non--chiral singlets will remain light, in which
case the two--step seesaw mechanism could not be phenomenologically
viable. This is in a sense reminscent of the supersymmetry $\mu$--problem.
The string models, however, also exhibits cases in which some, typically
undesirable states remain necessarily light. For example, in the string model
of ref. \cite{custodial} it is found that the right--handed neutrino are 
necessarily light in that model due to a local discrete symmetry.
Similarly, it is has also been observed that some string models
contain exotic fractionally charged states that cannot decouple
from the massless spectrum. This suggests that 
there exist string vacua in which the mass of some of the Standard Model 
singlets with $U(1)_{Z^\prime}$ charges is protected by a local
discrete symmetry. These fields will play the role of the
$SO(10)$ singlets in the seesaw mass matrix.

To study these aspects we study in some detail
an explicit supersymmetric solution.
The cubic level superpotential and the anomalous as
well as the anomaly free,
$U(1)$ combinations are given in ref. \cite{eu}. 
As an example, we find a solution to the 
$F$ and $D$ cubic level flatness constraints
with the following set of fields
\begin{equation}
\{{\bar V}_2, {V}_3,
H_{18}, H_{23}, H_{25},
\Phi_{45}, 
{\bar\Phi}_1^-,\Phi_2^+,{\bar\Phi}_3^-,
{\bar\Phi}_{23},
{\bar\Phi}_{13}, 
\xi_1\},
\label{firstsolution}
\end{equation}
having non--zero VEVs and all other fields have vanishing VEV. 
With this set 
of fields the general solution is 
\begin{eqnarray}
&&\vsq{H_{23}}~=~\h18-\p23-{1\over6}\v32\label{h23}\\
&&\vsq{H_{25}}~=~\p23+{1\over6}\v32\label{h25}\\
&&\vsq{\Phi_{45}}~=~3\anomaly+\h18-{1\over{10}}\v32\label{p45}\\
&&\vsq{{\bar\Phi_{13}}}~=~\anomaly-{1\over5}\v32\label{p13}\\
&&\vsq{\Phi^+_2}~=~\anomaly-{8\over{15}}\v32\label{p2plus}\\
&&\vsq{{\bar\Phi}_3^-}~=~\anomaly-{1\over{30}}\v32\label{p3minus}\\
&&\vsq{{\bar\Phi}_1^-}~=~\anomaly-{8\over{15}}\v32\label{p1minus}\\
&&\vsq{{\bar V}_2}~=~\v32\label{v2}\\
&&\l{\xi_1}\r~=~-{{\l{\bar\Phi}_{23}\r\l{H_{25}}\r}
			\over{\l{H_{23}}\r}}\label{xsi}
\end{eqnarray}

In this solution the VEVs of three fields, $\{ V_3,
{\bar\Phi}_{23},H_{18}\}$ remain as
free parameters, which are restricted to give a positive
definite solution for the set of $D$--term equations.

We start with $\{ V_3,{\bar\Phi}_{23},H_{25}\}=0$. 
In this case the set of fields with non--vanisnihing VEVs at
the string scale contains,
\begin{equation}
\{
H_{18}, H_{23},
\Phi_{45}, 
{\bar\Phi}_1^-,\Phi_2^+,{\bar\Phi}_3^-,
{\bar\Phi}_{13}
\}.
\label{secondsolution}
\end{equation}
the cubic level terms $h_1{\bar h}_3{\bar\Phi}_{23}+
{\bar h}_2H_{15}H_{18}$ give heavy mass to ${\bar h}^{2,3}$ and
the light Higgs representation contains $\bar h_1$ and ${\bar h}_{45}$. 
At $N=3$ we therefore have one Dirac neutrino mass term
\beq
\lambda_tU_1Q_1{\bar h}_1+\lambda_NN_1L_1{\bar h}_1
\label{neq3dmt}
\eeq
with $\lambda_t=\lambda_N$. There are no Dirac
mass terms at $N=4$; only a suppressed correction
to (\ref{neq3dmt}) at $N=5$, and none at $N=6$.
At $N=7$ we get
\beqn
&N_3L_1{\bar h}_1\Phi_{45}{\bar\Phi}_3^-T_1{\bar T}_3\nonumber\\
&N_3L_2{\bar h}_1\Phi_{45}{\bar\Phi}_3^-T_2{\bar T}_3\nonumber\\
&N_1L_2{\bar h}_1\Phi_{45}{\bar\Phi}_1^-T_2{\bar 1}_3\nonumber\\
&N_1L_2{\bar h}_1\Phi_{45}{\bar\Phi}_3^-T_1{\bar T}_2\nonumber\\
&N_1L_3{\bar h}_1\Phi_{45}{\bar\Phi}_3^-T_1{\bar T}_3\label{neq6dirac}
\eeqn
At $N=8$ there are only corrections to $N=3$. Similarly,
all $N=9$ order terms are suppressed compared to lower
order terms by six orders and there are new non--vanishing elements
in the neutrino Dirac mass matrix. At $N=10$ we have the new non--vanishing
element$$N_3L_3{\bar h}_1\Phi_{45}\Phi_{45}H_{18}H_{23}
{\bar\Phi}_3^-T_3{\bar T}_3.$$
With additional VEVs turned on at the string scale, more entries
in the neutrino Dirac mass matrix will be non--zero. In general we 
expect that the models retain some of the underlying $SO(10)$ symmetry
structure, and that the neutrino Dirac mass matrix is related
to the up quark mass matrix. Next we analyze the 
$N\phi$ mixing terms. Due to the large set of Standard Model 
singlets that may a priori couple to the $N_i$ fields, we follow
the following strategy. First we make a search up to $N=10$ 
to determine the set of fields that mix with the right handed
neutrinos. We then eliminate those that receive 
mass from cubic level terms in the particular vaccum solution.
We then determine the seesaw terms $N\phi$ with the set of
remaining massless fields. We assume here that the hidden sector
$SU(5)$ gauge group confines at $\Lambda\sim 10^{15}{\rm GeV}$
and that the ${\bar 5}5$ combinations form condensates of the 
hidden $SU(5)$ gauge group. We also take $\langle\phi\rangle/M\sim0.1$
for the set of fields with non--vanishing VEVs. 
With these assumptions the $N\phi$ mixing terms are: 
\beqn
N=6 &&N_2\Phi_2^+\Phi_{45}H_{23}H_{19}{\bar T}_2~~~~~~~~~~~\rightarrow 
10^{10}{\rm GeV} (N_2\phi_2^+;N_2\Phi_{45};N_2H_{23})\\
N=7 &&N_3H_{25}{\bar\Phi}_{13}\Phi_{45}{\bar\Phi}_3^-H_{19}{\bar T}_3~~~~~~
\rightarrow 10^9{\rm GeV} (N_3H_{45})\\
N=8 &&N_2\Phi_{13}{\bar\Phi}_{13}\Phi_2^+\Phi_{45}H_{23}
{\bar T}_2H_{19}~\rightarrow~ 10^{8}{\rm GeV}(N_2{\Phi}_{13})\\
    &&N_2{\bar\Phi}_{2}^+{\Phi}_{2}^+\Phi_2^+\Phi_{45}H_{23}
{\bar T}_2H_{19}~~\rightarrow~ 10^{8}{\rm GeV}(N_2{\bar\Phi}_2^+)\\
    &&N_2{\Phi}_{3}^-{\bar\Phi}_{3}^-\Phi_2^+\Phi_{45}H_{23}
{\bar T}_2H_{19}~~\rightarrow~ 10^{8}{\rm GeV}(N_2{\bar\Phi}_2^+)\\
    &&N_2{\Phi}_{1}^-{\bar\Phi}_{1}^-\Phi_2^+\Phi_{45}H_{23}
{\bar T}_2H_{19}~~\rightarrow~ 10^{8}{\rm GeV}(N_2{\bar\Phi}_2^+)\\
N=10 &&N_1H_{25}H_{23}H_{18}{\bar\Phi}_{13}\Phi_{45}\Phi_{45}{\bar\phi}_1^-
{\bar T}_2H_{19}~~\rightarrow~ 10^{6}{\rm GeV}(N_1H_{25})~~~~~~
\eeqn
Next we analyze the singlet mixing terms $\phi_i\phi_j$. In the
supersymmetric vacuum of eq. (\ref{secondsolution})
there are no mixing terms at
orders $N=4,5,6$. At $N=7$ we get terms of the form $\phi_i\phi_j
(T{\bar T})\phi^3$ with the following terms appearing
\beqn
&(&
{H}_{23}(\Phi_{13}+\Phi_{45}+{\Phi}_1^-+\Phi_2^++{\bar\Phi}_2^++
\Phi_3^-)+\nonumber\\
& &{\Phi}_{45}(\Phi_{13}+\Phi_{45}+\Phi_1^-+\Phi_2^++  
{\bar\Phi}_2^++{\Phi}_3^-)+\Phi_{2}^+{\bar\Phi}_{2}^+)
(T{\bar T})\Phi^3+\nonumber\\
& &(\Phi_1^-+{\bar\Phi}_2^++\Phi_3^-)\Phi_{13}\phi^5
\eeqn
At $N=8$ we get
\beq
\Phi_{45}\Phi_{45}T{\bar T}\phi^4+
\Phi_{45}H_{23}T{\bar T}\phi^4
\eeq

At $N=9$ we get terms that already appear at higher orders suppressed
by additional $\langle VEV\rangle/M$. We only list here new terms
that are unsuppressed as compared to the lower order terms. These are
\beqn
( && \Phi_2^+\Phi_1^-+
 \Phi_2^+\Phi_3^-+
 \Phi_2^+\Phi_{13}+
 \Phi_1^-\Phi_1^-+
 \Phi_1^-\Phi_3^-+
 \Phi_1^-{\bar\Phi}_2^++\nonumber\\
 && ~~~~~~~~~~~~~~~~~~
 \Phi_3^-\Phi_3^-+ \Phi_3^-{\bar\Phi}_2^++
 {\bar\Phi}_2^+{\bar\Phi}_2^++
 \Phi_{13}\Phi_{13}~~~)T{\bar T}\phi^5+ \nonumber\\
&& ~~~~~~~~~~~~~~~~
( ~~~\Phi_{45}\Phi_{45}+H_{23}\Phi_{45}+H_{23}H_{23}~~~)\phi^7
\label{neq10phiphi}
\eeqn
All terms that appear at $N=10$ are suppressed compared to
lower order terms and no new terms appear. 
The resulting neutrino mass matric then takes the approximate form

\begin{equation}
{\bordermatrix{
         & L_3& L_2& L_1& & N_3& N_2& N_1& & H_{23}&H_{25}&\Phi_{13}&\Phi_{45}&{\bar\Phi}_1^-&{\bar\Phi}_3^-&\Phi_2^+&{\bar\Phi}_2^+\cr
L_3      &  0 &  0 &  0 & & 0  &  0 &  r & &   0   &  0   &   0     &      0  &     0        &       0      &   0    &     0        \cr
L_2      &  0 &  0 &  0 & & r  &  0 &  r & &   0   &  0   &   0     &      0  &     0        &       0      &   0    &     0       \cr 
L_1      &  0 &  0 &  0 & & r  &  0 &  v & &   0   &  0   &   0     &      0  &     0        &       0      &   0    &     0       \cr
         &    &    &    & &    &    &    & &       &      &         &         &              &              &        &             \cr
N_3      &  0 & r  &  r & & 0  &  0 &  0 & &   0   &  x   &   0     &      0  &     0        &       0      &   0    &     0        \cr
N_2      &  0 & 0  &  0 & & 0  &  0 &  0 & &   z   &  0   &   u     &      z  &      u       &       u      &   z    &     u        \cr
N_1      &  r & r  &  v & & 0  &  0 &  0 & &   0   &  w   &   0     &      0  &     0        &       0      &    0   &     0       \cr
         &    &    &    & &    &    &    & &       &      &         &         &              &              &        &             \cr
H_{23}   &  0 &  0 &  0 & & 0  &  z &  0 & &   p   &  0   &   x     &      p  &     x        &       x      &    x   &      x       \cr
H_{25}   &  0 &  0 &  0 & & x  &  0 &  w & &   0   &  0   &   0     &      0  &     0        &       0      &    0   &      0       \cr
\Phi_{13}&  0 &  0 &  0 & & 0  &  u &  0 & &   x   &  0   &   q     &      x  &     y        &       y      &    0   &      y       \cr
\Phi_{45}&  0 &  0 &  0 & & 0  &  z &  0 & &   p   &  0   &   x     &      p  &     x        &       x      &    x   &      x       \cr
\Phi_1^-&  0  &  0 &  0 & & 0  &  u &  0 & &   x   &  0   &   y     &      x  &     q        &       q      &    q   &      q       \cr
\Phi_3^-&  0  &  0 &  0 & & 0  &  u &  0 & &   x   &  0   &   y     &      x  &     q        &       q      &    q   &      q       \cr
\Phi_2^+&  0  &  0 &  0 & & 0  &  z &  0 & &   x   &  0   &   0     &      x  &     q        &       q      &    0   &      x       \cr
{\bar\Phi}_2^+
        &  0  &  0 &  0 & & 0  &  0 &  0 & &   x   &  0   &   y     &      x  &     q        &       q      &    x   &      q     \cr}}
\nonumber\\
\nonumber\\
\label{stringderivedseesaw}
\end{equation}
where 
\beqn
&& r\sim{10^{-6}}{\rm GeV}~~~v\sim10^2{\rm GeV}~~~
w\sim10^{6}{\rm GeV}~~~q\sim10^{7}{\rm GeV}~~~u\sim10^{8}{\rm GeV}~\nonumber\\
&& x\sim10^{9}{\rm GeV}~~~z\sim10^{10}{\rm GeV}~~~p\sim10^{11}{\rm GeV}~~~y\sim10^{13}{\rm GeV}~
\nonumber
\eeqn
We next proceed to examine the mass spectrum of (\ref{stringderivedseesaw}).
We emphasize, however, that our aim is to study qualitatively the possible
role of the $SO(10)$ singlet fields in generating the neutrino flavor
parameters, rather than to find a vaccum solution that 
produces a realistic neutrino spectrum. Indeed, the solution given
by eq. (\ref{firstsolution}) cannot produce realistic mixing
also in the quark sector \cite{ckm}.
Our aim is to demonstrate the complication of trying to extract 
useful information from the string models in regard to the 
neutrino spectrum. In this respect we note that the vacuum solution given by
(\ref{secondsolution}) is a highly simplified solution, which is adequate
for our purpose here. In this regard we note from Eq.
(\ref{stringderivedseesaw}) that the main difficulty in
implementing the ``two--step'' seesaw mechanism in the string 
theory is need to keep at least three of the $SO(10)$ singlet
fields light down to relatively low energies, whereas from
(\ref{stringderivedseesaw}) we note that the general expectation
is for for the additional singlets to get mass terms which
are at least comperable to their mass terms with the right--handed
neutrinos. We note, however, from the analysis of the nonerenomalizable
terms that the $SO(10)$ singlet mass terms are spread over several
energy scales. Thus, the mass eigenstates of (\ref{stringderivedseesaw})
produce several degenerate states with large mixing. The neutrino 
spectrum itself is, however, not realistic in this vacuum with the
lightest eigenstate being a $L_3$  that does not mix with the 
other states, and is of order $10{\rm eV}$.
We also note that in this vacuum the charm mass terms vanishes. 
A slightly more realistic spectrum can be obtained by turning
${\bar\phi}_{23}\ne0$, in which case at $N=5$ we have the Dirac mass term 
$N_2L_2{\bar h}_{45}\Phi_{45}{\bar\Phi}_{23}$.
The mass matrix (\ref{stringderivedseesaw}) then has the mass
eigenvalues $\{1.7\times10^{13},1.7\times10^{13},2\times10^{11},
1\times10^{10},9\times10^9,1\times10^9,1\times10^9,5\times10^6,
101,101,17.5,17.5,0.02,2.4^{-8}\}{\rm GeV}$. Thus, the lightest eigenvalue,
which is predominantly $L_3$, is of of order $10{\rm eV}$. 
The next lightest states are two nearly degenerate states of order
$17.5{\rm GeV}$ that contain $\sim70\%$ mixture of $\Phi_1^-$ and $\Phi_3^-$ 
with order $10\%$ mixing with $L_2$.  
The remaining spectrum is readily analyzed and contains 
mixtures of the right-handed neutrinos and the $SO(10)$ singlets. 
The detailed analysis is not particularly revealing, so we do not
elaborate on it here. The main lesson from our analysis
is the demonstration that although a simple and elegant 
reasoning for the neutrino spectrum can be motivated 
from string theory in the form of (\ref{mnuleft}),
obtaining it from string models is an entirely different
story. The main difficulty from the perspective of the 
string model construction is to understand how the 
singlet masses can be protected from being too massive.
Furthermore, the analysis is complicated due to the proliferation of
$SO(10)$ singlets in the massless string spectrum and the lack
of an apparent guiding symmetry. While the vacuum 
solution, eq. (\ref{secondsolution}) that we used for the
analysis is somewhat simple, it does illustrate the primary
difficulty in incorporating the two--step seesaw in the string models.
More complicated solutions will allow more detailed structure
for the neutrino Dirac mass matrices, and will allow more $SO(10)$
singlets to effectively mix with the right--handed neutrinos through
nonrenormalizable terms. At the same time, however, they will 
also generate more mass terms the $SO(10)$ singlets and hence the
primary difficulty remains. 
Finally, we also comment that, in general we may also try to 
implement the ``one--step'' seesaw in the string models
We also remark that another possibilty is to implement the 
``one--step''seesaw in the string models. In the class of 
models under consideration this necessitates the utilization of
the exotic $H$ fields that carry $-1/2 Q(Z^\prime)$ with respect to the
charge of the right--handed neutrino. The relevant terms are 
then of the form $NNHHHH\phi^n$. In the specific model
under investigation such terms were not found up to $N=14$
and hence cannot induce the seesaw mechanism. Another important
observation is that due the fact the right--handed neutrino
mass terms are obtained from nonrenormalizable terms, we cannot 
generically assume that the seesaw scale $m_\chi$ is of the 
order of the GUT or string scale, and the right--handed neutrino
majorana mass matrix can in general involve several disparate
scales. 

\section{Conclusions}

The neutrino mass and mixing spectrum that emerged over
the past few years adds to the flavor puzzle. The new twist, 
however, is the large mixing versus the small mixing that 
we anticipated from the quark sector and Grand Unification. 
In this paper we proposed that this disparity between
the quark and lepton sector can be readily understood
in the context of string theories due to the extensive
$SO(10)$ singlet spectrum that exists in these models. 
This leads to the ``two--step'' seesaw mechanism that 
indeed can easily account for the flavor parameters in the
neutrino sector. Implementing the seesaw mechanism in string
constructions poses a far greater challenge. Short of deriving
the seesaw mechanism from string models there are nevertheless
many interesting physics issues that the ``two--step'' seesaw
mechanism presents. Primarily, with respect to the
possibility that the new sterile states may be sufficiently light, 
{\it i.e.} of the order $10-100{\rm TeV}$, and produce
observable phenomenological or cosmological effects. 
We will return to these issues in future publications.

\section{Acknowledgments}

This work was supported in part by PPARC. 

\bibliographystyle{unsrt}

\begin{thebibliography}{99}

\bibitem{superk} Y. Fukuda \etal [Super-Kamiokande Collaboration],
\PRL{81}{1998}{1562}.
\bibitem{sno} Q.R. Ahmad \etal [SNO collaboration]
\PRL{89}{2002}{011301}.
\bibitem{kamland} K. Eguchi \etal [KamLAND collaboration],
\PRL{90}{2003}{021802}.
\bibitem{MSW} L. Wolfenstein, \PRD{17}{1978}{2369};\\
S.P. Mikheev and A.Y. Smirnov, 
{\it Sov.\ J.\ Nucl.\ Phys.}\/ {\bf 42} (1985) 913.
\bibitem{wmap} C.L. Bennett \etal, astro-ph/0302207;\\
	       D.N. Spergel \etal, astro-ph/0302209.

\bibitem{galaxy} O. Elgaroy \etal, \PRL{89}{2002}{38};\\
		T.J. Pearson \etal, astro-ph/0205388;\\
		C.I. Kuo \etal, astro-ph/0212289.
\bibitem{bajc} B. Bajc, G. Senjanovic and F. Vissani, \PRL{90}{2003}{051802}.

\bibitem{dmr} K.R. Dienes and J. March--Russell, \NPB{479}{1996}{113}.
\bibitem{ps}     A.E. Faraggi, \NPB{428}{1994}{111}; \PLB{520}{2001}{337}.

\bibitem{seesaw} M. Gell--Mann, P.Ramond and R. Slansky, in
{\it Supergravity}, ed. by D. Freedman and P. Van--Nieuwenhuizen
(North--Holland 1979), p. 315;\\
 T. Yanagida, Proc. Workshop
on ``Unified Theories and Baryon Number of the Universe'', eds.
Sawata and A. Sugamoto, KEK, Japan (1979); \\
R.N. Mohapatra and G. Senjanovic, \PRL{44}{80}{912}.

\bibitem{king} See {\it e.g.} S.F. King, hep-ph/0208270,
and references therein.

\bibitem{chooz} M. Apollonio \etal [CHOOZ Collaboration],
\PLB{466}{1999}{415}.

\bibitem{mnuclasses} See {\it e.g.} S.F. King, hep-ph/0208266,
and references therein.

\bibitem{revamp} I. Antoniadis, J. Ellis, J. Hagelin, and D.V. Nanopoulos,
                  \PLB{231}{1989}{65};\\
		J. Lopez, D.V. Nanopoulos, and K. Yuan, \NPB{399}{1993}{654}.

\bibitem{slm}	A.E. Faraggi, D.V. Nanopoulos, and K. Yuan,
				\NPB{335}{1990}{347};\\
		A.E. Faraggi, \PLB{274}{1992}{47}; \NPB{387}{1992}{239}; \\
		G.B. Cleaver \etal, \PLB{455}{1999}{135};
				\IJMP{16}{2001}{425};
                                \NPB{593}{2001}{471};
                                \MODA{15}{2000}{1191};
                                \IJMP{16}{2001}{3565};
                                \NPB{620}{2002}{259}.

\bibitem{alr} 	I. Antoniadis, G.K. Leontaris, and J. Rizos,
						\PLB{245}{1990}{161}.

\bibitem{eu} \AEF, \PLB{278}{1992}{131}.

\bibitem{ks} A. Kagan and S. Samuel, \PLB{284}{1992}{289}.

\bibitem{chl} S. Chaudhuri, G. Hockney, and J. Lykken, 
					\NPB{469}{1996}{357}. 

\bibitem{lrs} G.B. Cleaver, \AEF~and C. Savage, \PRD{63}{2001}{066001};\\
	      G.B. Cleaver, D.J. Clements and \AEF, \PRD{65}{2002}{106003}.

\bibitem{su421} G.B. Cleaver, \AEF~and S. Nooij, hep-ph/0301037.
 
\bibitem{fff} H. Kawai, D.C. Lewellen, and S.H.-H. Tye, \NPB{288}{1987}{1};\\
               I. Antoniadis, C. Bachas, and C. Kounnas,
						\NPB{289}{1987}{87}.

\bibitem{nahe} A.E. Faraggi and D.V. Nanopoulos, \PRD{48}{1993}{3288};\\
	       A.E. Faraggi, \IJMP{14}{1999}{1663}.

\bibitem{lola} J. Ellis, G.K. Leontaris, S. Lola and D.V. Nanopoulos,
{\it Eur.\ Phys.\ J.}\/ {\bf C9} (1999) 389.

\bibitem{ccf}	\AEF, \PRD{46}{1992}{3204};\\
S. Chang, C. Coriano and A.E. Faraggi, \NPB{477}{1996}{65};\\
C. Coriano, \AEF~and M. Pl\"umacher, \NPB{614}{2001}{233}.

\bibitem{fh} A.E. Faraggi and E. Halyo, \PLB{307}{1993}{311}.

\bibitem{hnm} A.E. Faraggi and J. Pati, \PLB{400}{1997}{314}.

\bibitem{kln} S. Kalara, J.L. Lopez and D.V. Nanopoulos, 
\NPB{353}{1991}{650}.
\bibitem{dsw} M. Dine, N. Seiberg and E. Witten, \NPB{289}{1987}{585}.

\bibitem{fermionmasses}
J.L. Lopez and D.V. Nanopoulos, \NPB{338}{1990}{73}; \PLB{251}{1990}{73};
 \PLB{251}{1990}{73}; \PLB{268}{1991}{359};\\
J. Rizos and K. Tamvakis, \PLB{251}{1990}{369};\\
I. Antoniadis, J. Rizos and K. Tamvakis, \PLB{278}{1992}{257};
 \PLB{279}{1992}{281};\\
G.B. Cleaver \emph{et al.}, \PRD{57}{1998}{2701}; \PRD{59}{1999}{055005}.

\bibitem{ckm} A.E. Faraggi, \NPB{403}{1993}{101}; \NPB{407}{1993}{57};
		\PLB{329}{1994}{208};\\
	A.E. Faraggi and E. Halyo, \PLB{307}{1993}{305}; \NPB{416}{1994}{63}.
		
\bibitem{otherfermionmasses}
J.A. Casas and C. Munoz \NPB{332}{1990}{189};\\
J. Giedt, \NPB{595}{2001}{3};\\
S.A. Abel and C. Munoz, JHEP{0302}{2003}{010}.

\bibitem{rnm} R.N. Mohapatra and J.W.F. Valle, \PRD{34}{1986}{3457};\\
  J. Ellis, J. Hagelin, S. Kelley and D.V. Nanopolous, \NPB{311}{1988}{1};\\
  \AEF, \PLB{245}{1990}{435}.

\bibitem{leptogenesis} W. Buchm\"uller and T. Yanagida, 
	\PLB{302}{1993}{240};\\
  S. Davidson and A. Ibarra, \NPB{535}{2002}{25};\\
  W. Buchm\"uller, P. di Bari and M. Pl\"umacher, \NPB{643}{2002}{367};\\
  J. Ellis and M. Raidal, \NPB{643}{2002}{229}.

\bibitem{thor} M. Thormeier and A.E. Faraggi, \NPB{624}{2002}{163}.

\bibitem{custodial}     A.E. Faraggi, \PLB{339}{1994}{223}.


\end{thebibliography}

\vfill
\eject

\end{document}